\documentclass[11pt,tightenlines,onecolumn, aps, prd, nofootinbib, superscriptaddress, showkeys, showpacs]{revtex4-2}

\usepackage{latexsym}
\usepackage{amssymb}
\usepackage[dvips]{graphicx}
\usepackage{color}
\usepackage{graphicx}
\usepackage{caption}
\usepackage{subcaption}
\usepackage{todonotes}

\usepackage{amsmath, amsfonts, amssymb, mathrsfs}
\usepackage{amsthm}
\usepackage{tensor}
\usepackage{multirow}
\usepackage{bbm}
\usepackage[geometry]{ifsym}

\usepackage{float}

\newcommand{\Dslash}{{D\kern-0.63em{/}}}

\begin{document}

\newcommand{\beq}{\begin{equation}}
\newcommand{\eeq}{\end{equation}}

\title{Friedmann's Universe Controlled by Gauss-Bonnet Modified Gravity}

\author{F. dos Anjos}
\email{fabio.anjos@protonmail.com}

\affiliation{ Centro de Estudos Avan\c{c}ados de Cosmologia (CEAC/CBPF) \\
 Rua Dr. Xavier Sigaud, 150, CEP 22290-180, Rio de Janeiro, Brazil. }

\author{M. Novello}

\affiliation{ Centro de Estudos Avan\c{c}ados de Cosmologia (CEAC/CBPF) \\
 Rua Dr. Xavier Sigaud, 150, CEP 22290-180, Rio de Janeiro, Brazil. }

\date{\today}
\begin{abstract}
    We consider a Lagrangian to describe gravity using a nonlinear term depending on the Gauss-Bonnet invariant. We examine the conditions for a bouncing and the existence of an ulterior accelerated phase of the Universe.
\end{abstract}

\vskip2pc
\maketitle

\section{Introduction}

The accepted idea that the expansion of the universe is accelerating needs, for compatibility to general relativity, the introduction of some unusual forms of matter. However, several authors have proposed that instead of making weird hypothesis on some yet unobservable species of matter, one should follow the original idea of the first Einstein\rq s paper on cosmology and consider that in the cosmic scene one has to modify the equations that control the gravitational metric. This possibility led us to re-examine the evolution of the topological invariant containing two duals in a dynamical universe, the so called Gauss-Bonnet topological invariant. The particular interest on this invariant is due to the fact that in a homogeneous and isotropic universe this invariant drives the cosmic acceleration. In a decelerating scenario and as a necessary previous condition of an ulterior acceleration this invariant must have an extremum identified to its maximum value. We will examine the conditions for this to occur, and a description of the universe with epochs of accelerated and decelerated expansion.

Modified theories of gravity have been proposed in order to account for astronomical data, like the cosmic late acceleration. Two of such propositions that have been gaining notoriety are the f(R) theory and the Gauss-Bonnet (GB) topological invariant modified gravity \cite{lovelock1971, odintsov2005, odintsov2007, demartino2020}. In GB modified gravity it makes use of the topological invariant of the same name, which is only an invariant in 4D space-time, commonly expressed as $\mathcal{G}$. 

This topological invariant $\mathcal{G}$ by itself does not affect the dynamics of General Relativity in 4D when added to the Einstein-Hilbert action, instead, there are three basic ways for implementing it. First, by coupling $\mathcal{G}$ to a scalar field $\phi$ \cite{uddin2009,odintsov2019a, kawai2021}. Second, by considering a nontrivial function of the topological invariant, $f(\mathcal{G})$ modified gravity. And third, by implementing the topological invariant to a Einstein-Hilbert action of higher dimension $D$, in which $\mathcal{G}$ is no longer a topological invariant, and taking the limit $D\rightarrow4$, yielding nontrivial modified dynamics \cite{glavan2020}. Some groups had used observational data in order to find constraints to different approaches to GB modified gravity \cite{Benetti_2018, molavi2019, ODINTSOV2019134874, Camci_2021, kawai2021a}. Great reviews of modified theories of gravity can be found at \cite{odintsov2023,Nojiri_2017, MYRZAKULOV_2013, Dialektopoulos_2018, Fernandes_2022}.

Studies in $f(\mathcal{G})$ gravity have shown its capabilities of mimicking the $\Lambda$CDM cosmology and describe radiation/matter epochs followed by cosmic acceleration \cite{defelice2009}. The main critique for $f(\mathcal{G})$ is due to the appearance of higher order derivative terms in the equations, which may lead to instabilities in the theory in first order perturbation models \cite{defelice2010, defelice2009a}. This requires the establishment of criteria and conditions to exclude possible unphysical solutions (see \cite{uddin2009} for example).        

In this present work we use the $f(\mathcal{G})$ Gauss-Bonnet Modified Gravity approach, by applying the simplest non-trivial function of $\mathcal{G}$, we peer into its consequences on a homogeneous and isotropic Friedmann's geometry, first in an empty space-time, then in the case with source matter as cosmic dust. \textit{Section II.} gives a very brief definition of topological invariants in 4D. \textit{Section III.} presents the standard $f(\mathcal{G})$ gravity; \textit{subsection III.a.} presents a relationship between $f(\mathcal{G})$ and the Fierz representation of a spin-2 particle. \textit{Section IV.} applies the theory to the Friedmann's universe; \textit{subsection IV.A.} deals with the case for $\mathcal{G}=constant$, which yields a number of important constraints to filter out non-physical solutions; \textit{subsection IV.B.} considers the simplest non-trivial function of $\mathcal{G}$, that is $f(\mathcal{G})=\sigma \mathcal{G}^2$; first we assume an empty space-time, which produces a 2D dynamical system, and a complete qualitative description of numerical solutions near and around the critical points is presented; and second, we assume cosmic dust as source, producing a 3D dynamical system, and present a set of qualitative behavior of numerical solutions. \textit{Section V}, the conclusion. 

\section{Topological Invariants in 4D} 

 There are two topological invariants in a 4-d Riemannian geometry, that is

\begin{equation}
    I = \int \sqrt{- g} \, A \, d^{4}x,
\end{equation}

\begin{equation}
    II = \int \sqrt{- g} \, \mathcal{G} \, d^{4}x,
\end{equation}   
 where
\begin{equation}
    A = R_{\alpha\beta\mu\nu}^{*} \, R^{\alpha\beta\mu\nu}
\end{equation}    
and $\mathcal{G}$ is the Gauss-bonnet topological invariant

\begin{equation}
    \mathcal{G} = R_{\alpha\beta}^{*}{}_{\mu\nu}^{*} \, R^{\alpha\beta\mu\nu}.
\end{equation}

The dual is defined for any anti-symmetric tensor $F_{\mu\nu}$ as

\begin{equation} 
    F_{\mu\nu}^{*} = \frac{1}{2} \, \eta_{\mu\nu\alpha\beta} \, F^{\alpha\beta},
\end{equation}
\begin{equation}
    \eta_{\mu\nu\alpha\beta} = \sqrt{-g} \, \epsilon_{\mu\nu\alpha\beta},
\end{equation}
where $ \epsilon_{\mu\nu\alpha\beta} $ is the completely anti-symmetric Levi-Civit\`a symbol.

We note that the invariant $ \mathcal{G} $ can be written in terms of the curvature tensor (without the dual operation)  by the identity
\begin{equation}
    \mathcal{G} = - R_{\alpha\beta\mu\nu} \, R^{\alpha\beta\mu\nu} + 4 \, R_{\mu\nu} \, R^{\mu\nu} - R^{2}.
\end{equation}
This $\mathcal{G}$ satisfies the identity,
\begin{equation}
    \tensor{R}{^{\overset{\ast}{\mu\nu}} ^{\overset{\ast}{\alpha\beta}}} \tensor{R}{^\epsilon_\nu_\alpha_\beta} = \frac{1}{4} \mathcal{G} \tensor{g}{^\mu^\epsilon}.
\end{equation}

\newpage

\section{Modified Gravity via Gauss-Bonnet Topological Invariant}

For an arbitrary function $f(\mathcal{G})$, the variation of the action
\begin{equation}\label{eqn:acao2}
	S=\int \sqrt{-g}\left(R+f(\mathcal{G})+L_m\right),
\end{equation}
where $L_m$ is the matter Lagrangian, leads to the following set of dynamic equations,
\begin{equation}\label{eqn:RG+Q}
	R_{\mu\nu}-\frac{1}{2}Rg_{\mu\nu}+Z_{\mu\nu}=-T_{\mu\nu},
\end{equation}
where $Z_{\mu\nu}$ is defined as
\begin{equation}\label{eqn:Zmunu}
	Z_{\mu\nu}\equiv \frac{1}{2}\left(f'\mathcal{G}-f\right)g_{\mu\nu}+2\tensor{H}{^\alpha_{\;(\mu\nu);\alpha}},
\end{equation}
\begin{equation}\label{eqn:tensorH}
	\tensor{H}{^\mu^\nu^\beta}\equiv f''\mathcal{G}_{,\sigma} \tensor{R}{^{\overset{\ast}{\mu\nu}} ^{\overset{\ast}{\beta\sigma}}},
\end{equation}
where $f'=\frac{df}{d\mathcal{G}}$.

Note that contraction of the last two indices of $\tensor{H}{^\mu^\nu^\alpha}$ implies the identity
\begin{equation}
    \tensor{H}{^\mu^\nu_\nu}= -f''\mathcal{G}_{,\nu} \left(\tensor{R}{^\mu^\nu}-\frac{1}{2}R\tensor{g}{^\mu^\nu}\right).
\end{equation}

\subsection{The Curious Presence of a Fierz Tensor $H_{\alpha\beta\mu}$ Representing a Spin-2}

Let us show that such quantity $\tensor{H}{^\mu^\nu^\beta}$ defined by equation (\ref{eqn:tensorH}) may be associated to a spin-2 field. We start by remembering that there are two basic representations to deal with a spin-2 field that we call the Einstein and the Fierz representations (both were introduced by Fierz):
\begin{itemize}
    \item Einstein representation: $\varphi_{\mu\nu},$
    \item Fierz representation: $F_{\alpha\beta\mu}.$
\end{itemize}
The tensor $\varphi_{\mu\nu}$ is symmetric and it has 10 independent components. The tensor $F_{\alpha\beta\mu}$ has the symmetries
\begin{equation}
    \tensor{F}{^\mu^\nu^\beta}=-\tensor{F}{^\nu^\mu^\beta},
\end{equation}
which states that  this tensor has 24 independent components. The second property states that the dual has no trace, that is,
\begin{equation}
    \tensor{F}{_\mu_\nu_\beta}+\tensor{F}{_\beta_\mu_\nu}+\tensor{F}{_\nu_\beta_\mu}=0,
\end{equation}
or, equivalently,
\begin{equation}
    \tensor{F}{^{\overset{\ast}{\mu\alpha}}_\alpha}=0,
\end{equation}
which reduces this number to 20. The identity
\begin{equation}
    \tensor{F}{^\mu^\nu^\beta_{;\beta}}=0
\end{equation}
makes this number to 14, and finally, the property of vanishing trace,
\begin{equation}
    \tensor{F}{^\alpha^\mu_\mu}=0,
\end{equation}
reduces the number of independent components to 10.

It is a direct exercise to show that the tensor $\tensor{H}{^\mu^\nu^\alpha}$, defined in equation (\ref{eqn:tensorH}),
\begin{equation}
    f''\mathcal{G}_{,\sigma} \tensor{R}{^{\overset{\ast}{\mu\nu}} ^{\overset{\ast}{\beta\sigma}}},
\end{equation}
satisfies all the conditions above, showing that there is a spin-2 tensor hidden in the above dynamics, constructed with the invariant $\mathcal{G}$. 

The tensor $\tensor{H}{^\mu^\nu^\beta}$ has four properties:
\begin{itemize}
    \item Anti-symmetric in the first two indices: $\tensor{H}{^\mu^\nu^\beta}=-\tensor{H}{^\nu^\mu^\beta}$

    \item Cyclic identity: $\tensor{H}{_\mu_\nu_\beta}+\tensor{H}{_\beta_\mu_\nu}+\tensor{H}{_\nu_\beta_\mu}=0 \Leftrightarrow \tensor{H}{^{\overset{\ast}{\mu\alpha}}_\alpha}=0$

    \item Null divergent in the last indice: $\tensor{H}{^\mu^\nu^\beta_{;\beta}}=0$

    \item Contraction of the last two indices implies $:\tensor{H}{^\mu^\nu_\nu}= -X_{\nu} \left(\tensor{R}{^\mu^\nu}-\frac{1}{2}R\tensor{g}{^\mu^\nu}\right)$
\end{itemize}
So the quantity $\tensor{H}{^\mu^\nu^\beta}$ satisfies all these conditions, showing that it could be associated to  a spin-2 tensor hidden in the above dynamics. This is only a formal property of the quantity $H^{\mu\nu\alpha}$ related to derivatives of second order of the metric tensor.

\section{Gauss-Bonnet Modified Gravity in Friedmann's Universe}

Let us consider the homogeneous and isotropic Friedmann's metric,
\begin{equation}
    ds^2=dt^2-a(t)^2(dx^2+dy^2+dz^2),
    \label{6321}
\end{equation}
where $a(t)$ is the scale factor. Using the definition of the Hubble parameter $H\equiv \dot{a}/a$, the Gauss-Bonnet topological invariant yields
\begin{equation}\label{eqn:Q_Friedmann}
    \mathcal{G}=-24\frac{\ddot{a}\dot{a}^2}{a^3}= -24(\dot{H}+H^2)H^2,
\end{equation}
note that the sign of $\mathcal{G}$ indicates the sign of the acceleration of the scale factor.

From (\ref{eqn:RG+Q}), the new term $Z_{\mu\nu}$ can be treated as a geometric source, and in case of the metric (\ref{6321}), when $ \mathcal{G} $ is only a function of time, it can be associated to a perfect fluid \cite{Capozziello_2019}, that is,
\begin{equation}\label{eqn:RG+Q_fluido_Z}
    	R_{\mu\nu}-\frac{1}{2}Rg_{\mu\nu}=-Z_{\mu\nu}-T_{\mu\nu},
\end{equation}
where $Z_{\mu\nu}$ can be written under the form
\begin{equation}
    Z_{\mu\nu}=(\rho_z+p_z)v_\mu v_\nu -p_z g_{\mu\nu},
\end{equation}
Thus, the corresponding geometric density $\rho_z$ and geometric pressure $p_z$ are given by,
\begin{equation}\label{eqn:0.01}
    \rho_z=\frac{1}{2}\left(f'\mathcal{G}-f\right)+12f'' \dot{\mathcal{G}} H^3
\end{equation}
and
\begin{equation}\label{eqn:0.02}
    	p_z=-\frac{1}{2}\left(f'\mathcal{G}-f\right)-4\left[\left(f'''\dot{\mathcal{G}}^2+f''\ddot{\mathcal{G}}\right)H^2+2f''\dot{\mathcal{G}} H(\dot{H}+H^2)\right].
\end{equation}

The modified dynamical equations from (\ref{eqn:RG+Q_fluido_Z}) are straightforward,
\begin{align}
    \label{eqn:0.1}
    3H^2  &= \rho_z+\rho_m,\\
    \label{eqn:0.2}
    2\dot{H}+3H^2 &= -p_z-p_m,
\end{align}
where the energy-matter source satisfies the conservation law,
\begin{equation}\label{eqn:0.3}
    0 =\dot{\rho_m}+3(\rho_m+p_m)H.
\end{equation}
Observe that, differentiating \eqref{eqn:0.1} with respect to time, and applying equations \eqref{eqn:Q_Friedmann} and \eqref{eqn:0.3}, results in \eqref{eqn:0.2}, that is, \eqref{eqn:0.2} is not an independent equation and can be interpreted as a  cinematic relation with respect to the expansion parameter of the model.

In order to observe qualitative and quantitative properties of the model we make use of the theory of dynamical systems. Choosing the variables $\mathcal{G}$, $H$ and $\rho_m$, and through equations \eqref{eqn:Q_Friedmann}, \eqref{eqn:0.01}, \eqref{eqn:0.1} and \eqref{eqn:0.3}, the problem can be enunciated in a general dynamical system form:
\begin{align}
    \label{eqn:0.4}
    \dot{\mathcal{G}} = & \frac{1}{12f''H^3}\left[3H^2 -\frac{1}{2}(f'\mathcal{G}-f) -\rho_m\right]\\
    \label{eqn:0.5}
    \dot{H} = & -\frac{\mathcal{G}}{24H^2}-H^2\\
    \label{eqn:0.6}
    \dot{\rho_m} = & -3(\rho_m +p_m)H,
\end{align}
where \eqref{eqn:0.6} refers to one class of matter-energy source, and when other independent sources of matter-energy are added to the system there are the corresponding source density dynamical variables and equations. In addition, it is necessary to define the equation of state relating $\rho_m$ and $p_m$, usually of the form $p=\omega \rho$, for a given value of $\omega$.

One class of functions for the topological invariant $\mathcal{G}$ to consider is of the type $f(\mathcal{G})=\sigma\mathcal{G}^n$, where $\sigma$ is a constant parameter and $n=2, 3, ...$, in order to give rise to nontrivial effects. The choice of a particular $f(\mathcal{G})$ only affects equation \eqref{eqn:0.4},
\begin{equation}
    \dot{\mathcal{G}}=\frac{1}{12\sigma n(n-1)\mathcal{G}^{n-2}H^3}\left[3H^2-\frac{n-1}{2}\sigma\mathcal{G}^n -\rho_m \right].
\end{equation}
It is apparent that for $n>2$ the dynamical system presents a singularity when $\mathcal{G}=0$ in the phase space, which may provide a barrier for the system to transition between positive and negative values of $\mathcal{G}$. Then, $n=2$ provides the simplest case in order for $\mathcal{G}$ to have the ability to change sign, and consequently, give rise to a model with epochs of accelerated and decelerated rates of expansion. We will see later that this system can have solutions where the Gauss-Bonnet topological invariant $\mathcal{G}$ is cyclically dampened, changing it's sign periodically, which could be used to describe a universe with stages of acceleration and deceleration, but first let us consider the case where $\mathcal{G}$ is constant for an interval or indefinitely.

\subsection{The Case with $\mathcal{G}=cte$}

The most simple behavior we could get from $\mathcal{G}$ is a constant $\alpha$, let
\begin{equation}
    \mathcal{G}\equiv \alpha,
\end{equation}
so $\dot{\mathcal{G}}=0$ and $\ddot{\mathcal{G}}=0$, then the geometric density and pressure \eqref{eqn:0.01} and \eqref{eqn:0.02} are constants too given by
\begin{equation}\label{eqn:rho_alpha}
    \rho_z=\frac{1}{2}\left(f'(\alpha)\alpha-f(\alpha)\right),
\end{equation}
\begin{equation}
    p_z=-\frac{1}{2}\left(f'(\alpha)\alpha-f(\alpha)\right).
\end{equation}

From the definition of $\mathcal{G}$ in \eqref{eqn:Q_Friedmann} the problem is to solve the integral,
\begin{equation}\label{eqn:0.7}
    \int \frac{-H^2}{H^4+\frac{\alpha}{24}}d H = t+k.
\end{equation}
There are three classes of solutions for \eqref{eqn:0.7}, depending if $\alpha$ is positive, negative or zero. And there is a fourth class of solutions by assuming $H$ as a constant.

\subsubsection{Solution for $\alpha$ positive}

\begin{figure}
     \centering
     \begin{subfigure}[b]{.47\textwidth}
         \centering
         \includegraphics[width=\textwidth]{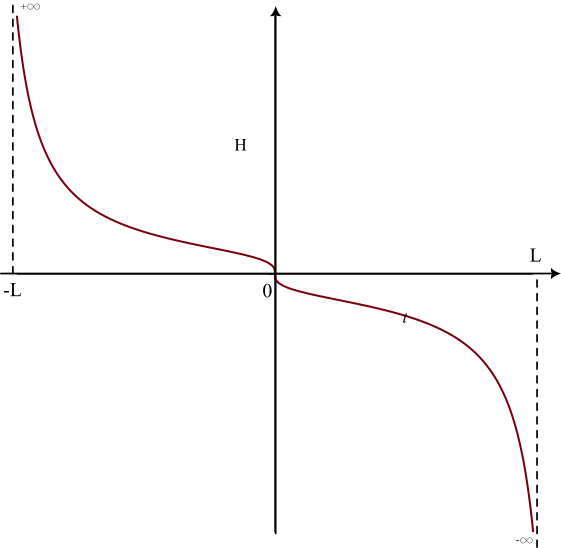}
         \caption{General solution of $H(t)$ for $\alpha>0$. The $H(t)$ diverges to $\pm\infty$ in a finite interval $[-L,L]$, where $L=\pm\frac{\pi}{2\sqrt{2}}\left(\frac{24}{\alpha}\right)^{1/4}$.}
         \label{fig:001}
     \end{subfigure}
     \hfill
     \begin{subfigure}[b]{.47\textwidth}
         \centering
         \includegraphics[width=\textwidth]{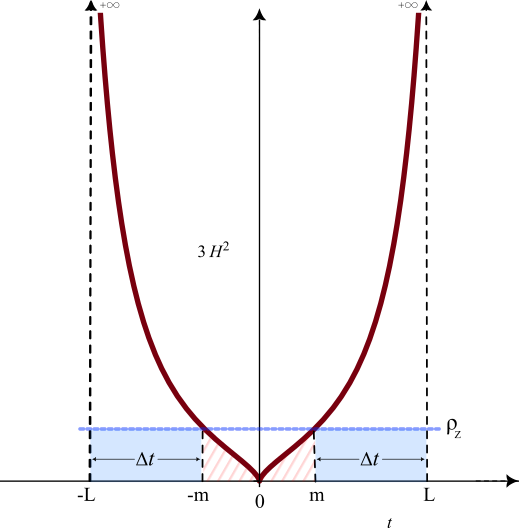}
         \caption{General graph of $3H^2$ as function of time. The $\Delta t$ delimits the intervals for which $\rho_m\geq 0$, the size of these intervals depend on the value of $\rho_z$. The function diverges to $+\infty$ at $\pm L$, and $m$ depends on the specific choice for $f(\mathcal{G})$ and the value of $\alpha$.}
         \label{fig:002}
     \end{subfigure}
     \caption{Solution of $H(t)$ for $\mathcal{G}$ as a positive constant.}
\end{figure}

The implicit solution of \eqref{eqn:0.7} for $\alpha>0$ is
\begin{equation}\label{eqn:0.8}
    \begin{split}
    & -\left(\frac{3}{8\alpha}\right)^{\frac{1}{4}} 
    \left[ 
        \frac{1}{2}\ln \left(
            \frac{H^2-\left(\frac{\alpha}{6}\right)^{1/4}H+\left(\frac{\alpha}{24}\right)^{1/2}}{H^2+\left(\frac{\alpha}{6}\right)^{1/4}H+\left(\frac{\alpha}{24}\right)^{1/2}}
        \right)
        +\arctan \left(
            2\left(\frac{6}{\alpha}\right)^{\frac{1}{4}}H+1
        \right)
        +\arctan \left(
        2\left(\frac{6}{\alpha}\right)^{\frac{1}{4}}H-1
        \right)
    \right] \\ 
    & = t+k,
    \end{split}
\end{equation}
where $k$ is a constant of integration that simply shifts the solution in time. Assuming $k=0$ and taking the limit for $H\rightarrow \pm\infty$ on the left side of \eqref{eqn:0.8}, it gives the result 
\begin{equation}
    L = \pm\frac{\pi}{2\sqrt{2}}\left(\frac{24}{\alpha}\right)^{\frac{1}{4}}.
\end{equation}
This implies that the Hubble parameter diverges to $\pm\infty$ in a finite interval $[-L,L]$. The general solution is expressed in \textit{figure \ref{fig:001}}.

From equation \eqref{eqn:0.1} the source density can be expressed as
\begin{equation}
    \rho_m=3H^2 -\rho_z,
\end{equation}
this means that if $\rho_z$ is positive, then the condition $\rho_m\geq 0$ is only satisfied in the intervals $-L<t\leq -m$ and $m\leq t<L$, where $\pm m$ is the time $t$ for which the relation $3H^2 -\rho_z =0$ is satisfied. The value of $m$ depends on the choice of the function $f(\mathcal{G})$ and $\alpha$. The conclusion is that, for $\mathcal{G}$ as a constant $\alpha$ and $\rho_z$ positive, the evolution of the Hubble parameter $H$ is confined in one of the intervals $[-L,-m]$ or $[L,m]$ in order for the source density $\rho_m$ to be greater or equal to zero, so $H$ can be either positive or negative in this configuration. See \textit{figure \ref{fig:002}} for the graph of $3H^2$ and its respective allowed intervals.

\subsubsection{Solution for $\alpha$ negative}

\begin{figure}
     \centering
     \begin{subfigure}[b]{.47\textwidth}
         \centering
         \includegraphics[width=\textwidth]{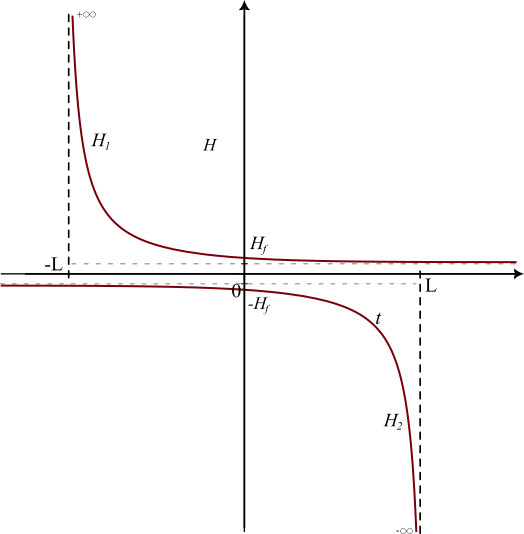}
         \caption{General solution of $H(t)$ for $\alpha<0$. The $H(t)$ has two possible solutions, $H_1$ positive and $H_2$ negative. The $H_1$ diverges to $+\infty$ at a finite past $t=-L$ (vertical dashed line), and approaches the value $H_f$ at the infinite future (horizontal dashed line). The $H_2$ is the reverse mirror image of $H_1$.}
         \label{fig:003}
     \end{subfigure}
     \hfill
     \begin{subfigure}[b]{.47\textwidth}
         \centering
         \includegraphics[width=\textwidth]{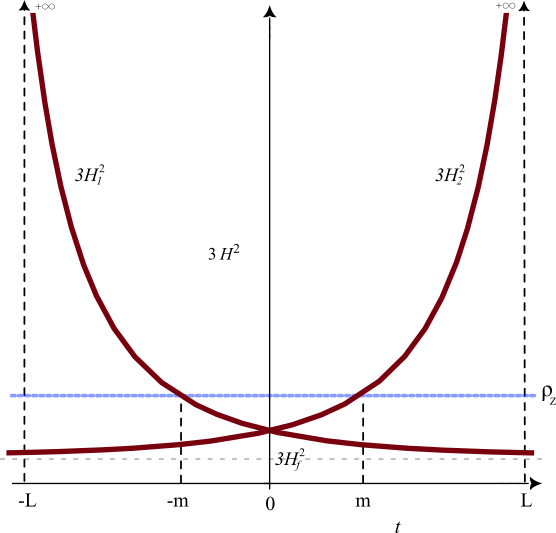}
         \caption{General graph of $3H_1^2$ and $3H_2^2$ as functions of time, both functions have a minimum value at $3H_f^2=3\sqrt{\frac{|\alpha|}{24}}$. If $3H_f^2 \geq\rho_z$ then $H_1$ and $H_2$ are allowed in their full respective time intervals. But if $3H_f^2\leq\rho_z$, then, in order for $\rho_m\geq 0$, the $H_1$ and $H_2$ are only allowed in the time intervals $[-L,-m]$ and $[m,L]$ respectively.}
         \label{fig:004}
     \end{subfigure}
     \caption{Solution of $H(t)$ for $\mathcal{G}$ as a negative constant.}
\end{figure}

The implicit solution of \eqref{eqn:0.7} for $\alpha<0$ is
\begin{equation}
    \left(\frac{3}{2|\alpha|}\right)^{\frac{1}{4}}
    \left[
        \frac{1}{2}\ln\left(\frac{H+\left(\frac{|\alpha|}{24}\right)^{\frac{1}{4}}}{H-\left(\frac{|\alpha|}{24}\right)^{\frac{1}{4}}}\right)
        -\arctan\left(\left(\frac{24}{|\alpha|}\right)^{\frac{1}{4}}H\right)
    \right]
    =t+k.
\end{equation}
Where $k$ is a constant of integration. For $k=0$, the implicit solution above implies two possible solutions for $H(t)$, one is $H_1 (t)$ which is always positive, it diverges to $+\infty$ at a finite past time $t=-L$, and for the infinite future ($t\rightarrow\infty$) it stabilizes at the value $H_f=\left(\frac{|\alpha|}{24}\right)^{1/4}$. The other solution is $H_2 (t)$ which is always negative, it diverges to $-\infty$ in a finite future $t=L$, and for the infinite past ($t\rightarrow -\infty$) it approaches the value $-H_f$. The solutions $H_1$ and $H_2$ are the inverse mirror images of each other, obtained by the transformation $H_2(t)=-H_1 (-t)$, see \textit{figure \ref{fig:003}}. The values of $\pm L$ is obtained by the limits $H_1\rightarrow +\infty$ and $H_2\rightarrow -\infty$, given by
\begin{equation}
    |L|=\sqrt{\frac{3}{2}}\frac{\pi}{(24|\alpha|)^{1/4}}.
\end{equation}

The functions $3H_1^2$ and $3H_2^2$ have the same minimum value at $3H_f^2$. Using the condition for the positivity of matter source ($\rho_m \geq 0$) given by \eqref{eqn:0.1}, that is, $3H^2\geq \rho_z$, it leads to two possibilities: (a) If $3H_f^2 \geq\rho_z$ then the solutions $H_1$ and $H_2$ are valid over their respective time intervals ($[-L,+\infty]$ and $[-\infty,L]$); (b) If $3H_f^2 \leq\rho_z$ then $H_1$ is only valid in the interval $[-L,-m]$, and the same for $H_2$ in the interval $[m,L]$, where $\mp m$ is the time for which $3H_1^2=\rho_z$ and $3H_2^2=\rho_z$ respectively, see \textit{figure \ref{fig:004}}. The value of $\pm m$ will depend on the specific choice of the function $f(\mathcal{G})$ and the value of $\alpha$.

\subsubsection{Solution for $\alpha=0$}

In the case for $\mathcal{G}$ as a constant with value zero, it is possible to obtain an explicit solution for the Hubble parameter. Solving \eqref{eqn:0.7} for $\alpha=0$ gives the solution for $H$,
\begin{equation}
    H=\frac{1}{t+K_1},
\end{equation}
where $K_1$ is a constant of integration. As for the scale factor $a(t)$
\begin{equation}
    a(t)=K_2(t+K_1),
\end{equation}
with $K_2$ a constant of integration.

So if $\mathcal{G}$ is constant and zero, then the scale factor increases or decreases linearly over time, and the Hubble parameter presents a singularity at $t=-K_1$. The condition for the positivity of the matter source ($\rho_m\geq 0$), can be obtained explicitly from equation \eqref{eqn:0.1}, 
\begin{equation}
    \label{eqn:4}
    \rho_m= \frac{3}{(t+K_1)^2}-\rho_{z}.
\end{equation}
From (\ref{eqn:4}) we see that if the geometric density $\rho_{z}$ is less or equal to zero, then $\rho_m$ is always positive; but if $\rho_{z}$ is positive, then $\rho_m$ will be positive only in the interval
\begin{equation}
    -\sqrt{\frac{3}{\rho_{z}}}-K_1\leq t\leq \sqrt{\frac{3}{\rho_{z}}}-K_1,
\end{equation}
which depends on the the initial condition for $H(t)$. 

\subsubsection{Solutions for $H$ as a constant}

And finally, there is one last class of solutions obtained by assuming the Hubble parameter as a constant $H_c$. For $H_c\neq 0$, and using the equation of state $p_m=\omega \rho_m$, the dynamical equation \eqref{eqn:0.1} and the conservation equation \eqref{eqn:0.3}, that is,
\begin{align}\label{eqn:4.1}
    \rho_m & =3H_c^2-\rho_z \\
    \label{eqn:4.2}
    \dot{\rho_m} & =-3\rho_m(1+\omega)H_c,
\end{align}
can only be satisfied if $\rho_m =0$, since \eqref{eqn:4.1} implies $\rho_m$ as a constant but \eqref{eqn:4.2} makes $\dot{\rho_m}\neq 0$ if $\rho_m\neq 0$. From \eqref{eqn:4.1} and \eqref{eqn:Q_Friedmann} we obtain the relationships between $H_c$, $\alpha$ and $\rho_z$,
\begin{equation}
    H_c^2=\frac{\rho_z}{3}=\sqrt{\frac{-\alpha}{24}},
\end{equation}
where $\alpha$ must be negative for the quantities to be real numbers. So a constant Hubble parameter can only describe an empty space-time.

It is worth noting that for $H_c = 0$ then from \eqref{eqn:Q_Friedmann} it implies $\alpha = 0$, and equations \eqref{eqn:4.1} and \eqref{eqn:4.2} are both satisfied, with the matter source assuming the condition,
\begin{equation}
    \rho_m = -\rho_z=\frac{1}{2}f(0),
\end{equation}
where \eqref{eqn:rho_alpha} was used. Here we have a description of a static universe with homogeneous matter density $\rho_m$ as a constant directly controlled by the function $f(0)$. Since the Hubble parameter is zero, this means the scale factor is a constant $a_c$ in the metric. The positivity of the matter source imposes the condition,
\begin{equation}
    \rho_m \geq 0 \Rightarrow f(0) \geq 0.
\end{equation}

In summary, when the topological invariant $\mathcal{G}$ is assumed to be constant, there emerges specific conditions in order for the matter source $\rho_m$ to be strictly positive, even in some cases the topological invariant $\mathcal{G}$ cannot be constant indefinitely but must vary over time. These specific conditions are highly dependent on the choice of the function $f(\mathcal{G})$. This may be useful in the future in order to filter only solutions with $\rho_m \geq 0$. The separation of solutions by the sign of $\mathcal{G}$ is useful because it indicates if the system is under accelerated, decelerated or null acceleration of expansion, according to \eqref{eqn:Q_Friedmann}. When the Hubble parameter is assumed constant it is clear the direct influence of the function $f(\mathcal{G})$ over the system, either dictating the expansion rate in an empty universe, or by controlling the value of the homogeneous source matter density $\rho_m$ in a static universe.

\subsection{Case with $f(\mathcal{G})=\sigma \mathcal{G}^2$}

We now consider the first non trivial term in the Taylor expansion of $f(\mathcal{G})$, that is, the function $f(\mathcal{G})=\sigma \mathcal{G}^2$, with $\sigma$ as a constant. In this case the geometric density and pressure assume the form,
\begin{align}
    \label{eqn:5.1}
    \frac{\rho_z}{\sigma}= & \frac{\mathcal{G}^2}{2}+24\dot{\mathcal{G}}H^3,\\
    \label{eqn:5.2}
    \frac{p_z}{\sigma}= & -\frac{\mathcal{G}^2}{2}-8\left[\ddot{\mathcal{G}}H^2+2\dot{\mathcal{G}}H\left(\dot{H}+H^2\right)\right],
\end{align}
and we have to solve the set of differential equations,
\begin{align}
    \label{eqn:6}
    3H^2 = & \rho_z+\rho_m,\\
    \label{eqn:8}
    0 = & \dot{\rho_m}+3(\rho_m+p_m)H,\\
    \label{eqn:9}
    \mathcal{G}= & -24(\dot{H}+H^2)H^2.
\end{align}

\subsubsection{Consider an Empty Space-Time}

Let us consider the case with an empty space-time. Under these conditions equation (\ref{eqn:8}) is identically zero. Applying the dynamical system expression of the problem, given by \eqref{eqn:0.4} and \eqref{eqn:0.5}, we have

\begin{align}
    \label{eqn:10}
    \dot{\mathcal{G}}= & \frac{1}{24H^3}\left[\frac{3H^2}{\sigma}-			\frac{\mathcal{G}^2}{2}\right],\\
    \label{eqn:11}
    \dot{H}= & -\frac{\mathcal{G}}{24H^2}-H^2.
\end{align}
This 2D dynamical system represents the whole problem in an empty space-time. Its numerical solutions will prove useful in giving a map to look for certain types of structures and behaviors in the complete problem with energy-matter sources, which is a dynamical system of higher dimension.

For $\dot{\mathcal{G}}=0$ and $\dot{H}=0$, the equilibrium points in the phase space $(\mathcal{G}(t) \times H(t))$ are $\mathcal{G}_0=-\frac{-1}{\sigma^{2/3}}\left(\frac{3}{2}\right)^{\frac{1}{3}}$ and $H_0=\pm\frac{1}{\left({\sigma 96}\right)^{1/6}}$, these imply the constant $\sigma$ must be positive in order for the equilibrium points to be real. We already have seen in subsection IV.A.4 that solutions for the Hubble parameter as a constant $H_0$ are viable in an empty space-time, which means that in these equilibrium points the scale factor behaves as an exponential manner, given by $a(t)\sim \exp{(H_0 t)}$. Observe that the line $H=0$ in the phase space is a singularity region, and for the origin $H=\mathcal{G}=0$ of the phase space, the dynamical system is undefined, but from \eqref{eqn:6} and \eqref{eqn:9} we see that the equations are identically satisfied. Then, in order to determine the behavior of solutions near the origin, specific methods of analyses must be applied. 

The next step is to determine the behavior of solutions near the equilibrium points. For this, we analyze the Jacobian matrix of the system in the equilibrium points, $a=\left(\frac{\partial\dot{\mathcal{G}}}{\partial \mathcal{G}}\right)_{(\mathcal{G}_0,H_0)}$, $b=\left(\frac{\partial\dot{\mathcal{G}}}{\partial H}\right)_{(\mathcal{G}_0,H_0)}$, $c=\left(\frac{\partial\dot{H}}{\partial \mathcal{G}}\right)_{(\mathcal{G}_0,H_0)}$ and $d=\left(\frac{\partial\dot{H}}{\partial H}\right)_{(\mathcal{G}_0,H_0)}$, and compute the eigenvalues of this matrix,
\begin{equation}
    \begin{vmatrix}
	a-\lambda & b\\
    c & d-\lambda
	\end{vmatrix}
	=0.
\end{equation}
It turns out that the two eigenvalues are real and of opposite signs for both equilibrium points, so these equilibrium points are saddle points, that is, trajectories near the equilibrium points are attracted or repelled depending on the direction they approach these points (see \cite{Bahamonde_2018}).

%\end{document}

\begin{figure}
    \centering
    \includegraphics[scale=0.5]{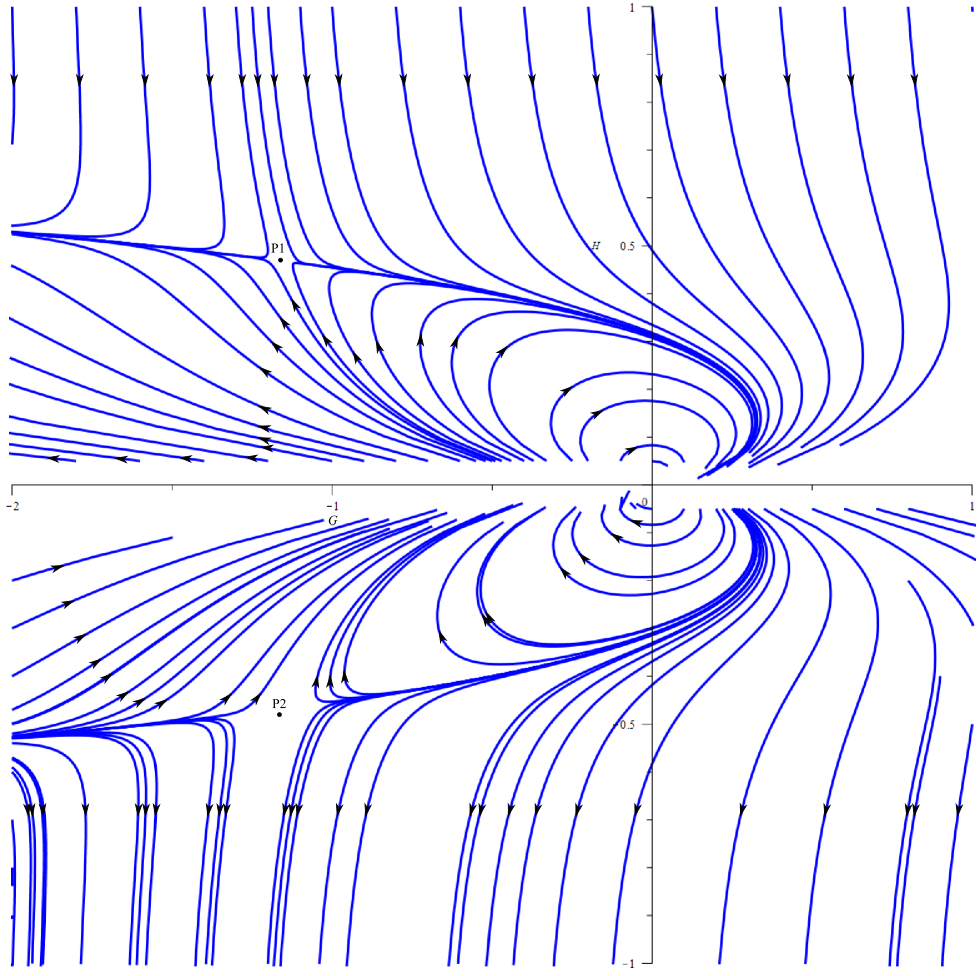}
    \caption{Integral solutions for the 2D dynamical system in the phase space $[\mathcal{G}(t)\times H(t)]$. The points $P1$ and $P2$ are the corresponding equilibrium points of the system. Observe that the $\mathcal{G}$ axis is a singularity for the system and we cannot infer their behavior numerically.}
    \label{fig:fig_01}
\end{figure}

Now to gain more information on the solutions for this problem we can numerically integrate the solutions for given initial conditions. To numerically integrate the solutions we assumed $\sigma\equiv 1$, and using the Runge–Kutta method (\cite{gould}), integrating for different initial conditions inside the interval $[\mathcal{G}=-2...1, H=-1...1]$ which contains the two equilibrium saddle points, gives the behavior map in \textit{figure \ref{fig:fig_01}}.

\begin{figure}
     \centering
     \begin{subfigure}[b]{.48\textwidth}
         \centering
         \includegraphics[width=\textwidth]{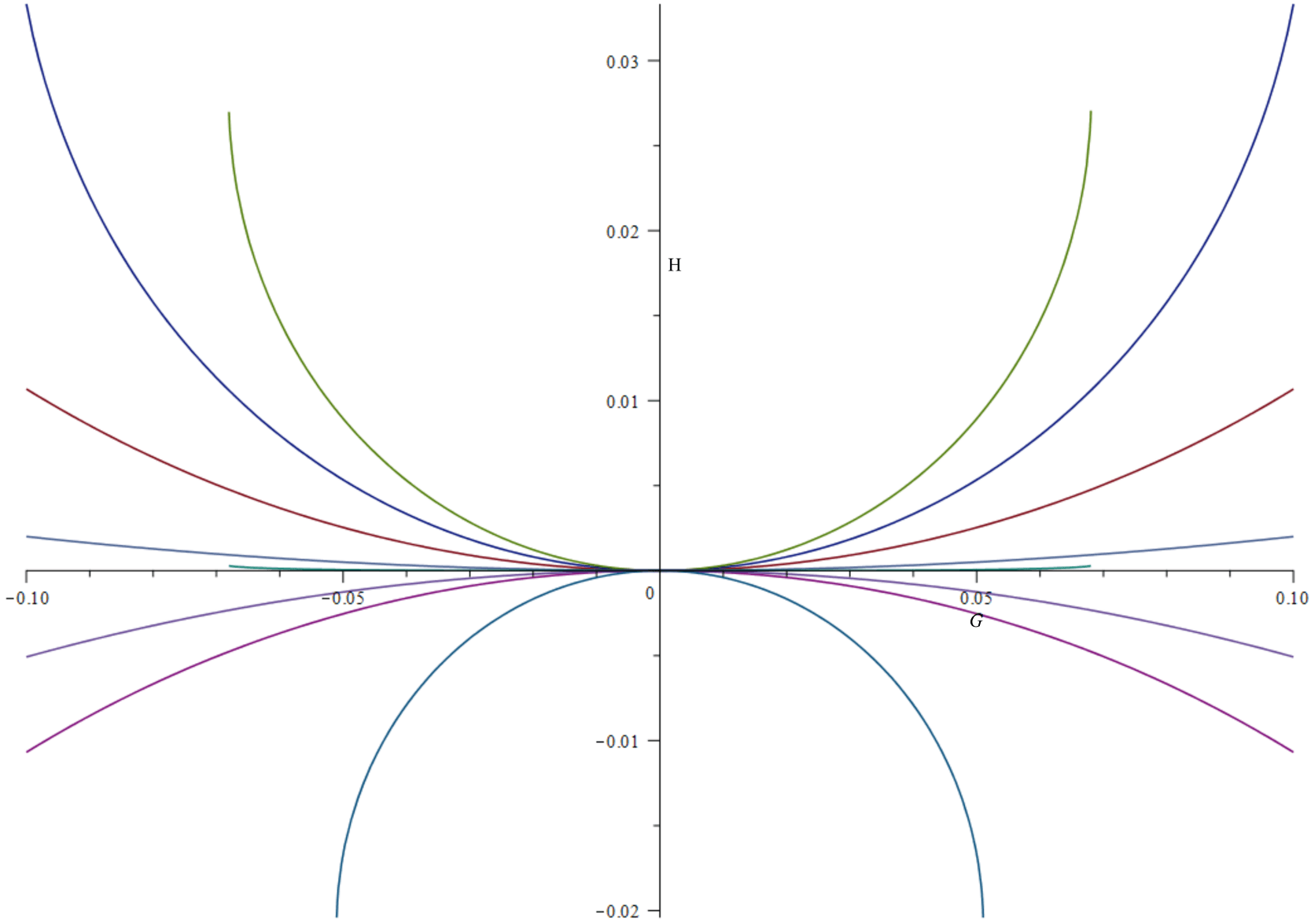}
         \caption{Plots of the solution (\ref{eqn:14}) for different initial conditions.}
         \label{fig:fig_02}
     \end{subfigure}
     \hfill
     \begin{subfigure}[b]{.46\textwidth}
         \centering
         \includegraphics[width=0.9\textwidth]{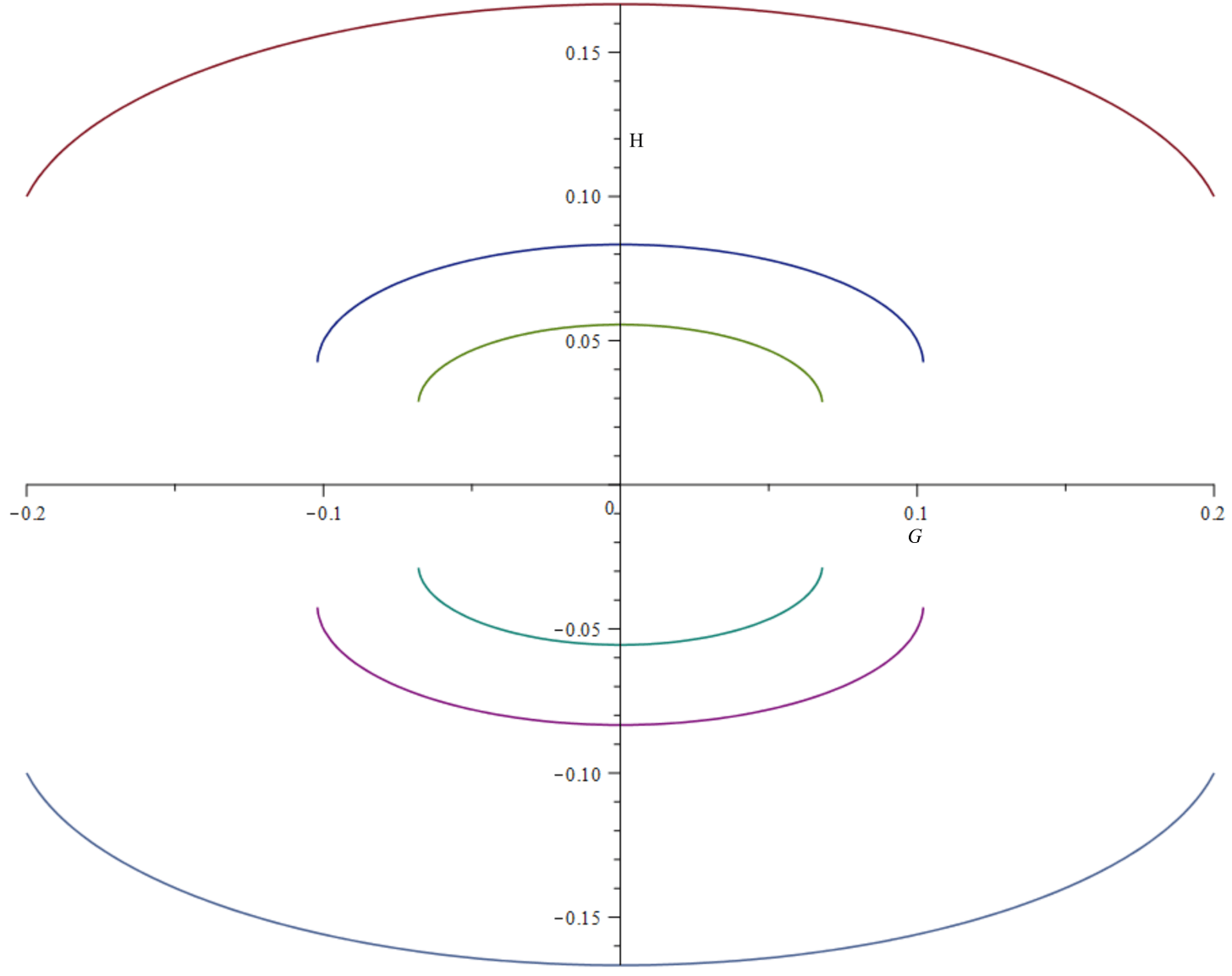}
         \caption{Plots of the solution (\ref{eqn:15}) for different initial conditions.}
         \label{fig:fig_03}
     \end{subfigure}
     \caption{Solutions near the origin of the phase space $\mathcal{G} \times H$.}
\end{figure}

Numerical solutions are only possible if not near the region of singularity of the system, but through approximation it is possible to determine the behavior of solutions near the singularity region (for $H \approx 0$). From (\ref{eqn:10}) and (\ref{eqn:11}) it follows,
\begin{equation}
    \label{eqn:12}
    \frac{\dot{H}}{\dot{\mathcal{G}}}= \frac{dH}{d\mathcal{G}}=  -H\frac{\left(\mathcal{G}+24H^4\right)}{3H^2-\frac{\mathcal{G}^2}{2}}.
\end{equation}
For $|H|<<1$ and $|\mathcal{G}|>>1$, equation (\ref{eqn:12}) simplifies to $\frac{dH}{d\mathcal{G}}\approx 0$, with solution $H(t)\approx cte $, that is, in the phase space the $H(t)$ is constant near the $\mathcal{G}$ axis when not close to the origin. Let us focus now on what happens to solutions near the origin of the phase space, since $\mathcal{G}\approx 0$ and $H\approx 0$, the $H^4$ term can be dropped out compared to $\mathcal{G}$ in equation (\ref{eqn:12}), yielding
\begin{equation}
    \label{eqn:13}
    \frac{dH}{d\mathcal{G}}\approx -\frac{H\mathcal{G}}{3H^2-\frac{\mathcal{G}^2}{2}},
\end{equation}
and the solutions to this system are given by,
\begin{equation}
    \label{eqn:14}
    	H(\mathcal{G})=\frac{1}{12a}\left[1-\sqrt{1-24a^2\mathcal{G}^2}\right],
\end{equation}
and
\begin{equation}
    \label{eqn:15}
    H(\mathcal{G})=\frac{1}{12b}\left[1+\sqrt{1-24b^2\mathcal{G}^2}\right],
\end{equation}
where $a$ and $b$ are constants that depend on the initial conditions. We note that the value of $H(t)$ is bounded. The plots of (\ref{eqn:14}) and (\ref{eqn:15}), for different values for $a$ and $b$, are given in \textit{figures \ref{fig:fig_02} and \ref{fig:fig_03}}. Although the dynamical system is undefined at the origin of the phase space, we see that its solutions near it are well defined, the origin acts as a convergence and divergence of solutions, depending on the region it is approached.      

\begin{figure}[h]
    \centering
    \includegraphics[scale=0.33]{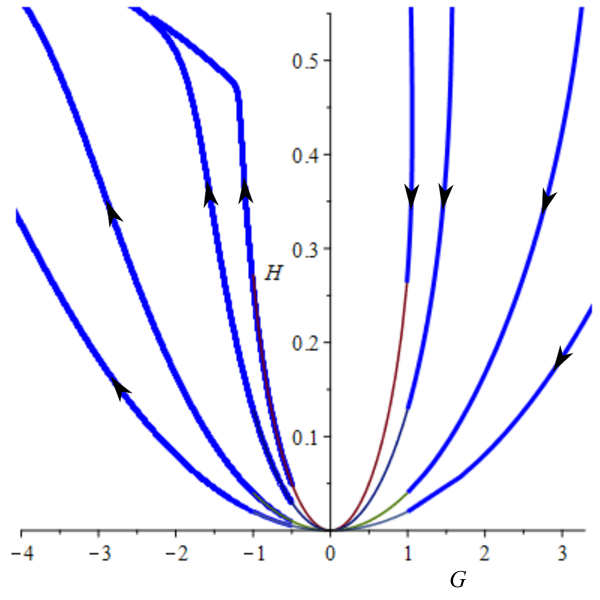}
    \caption{Examples of integral solutions that pass one time through the origin of the phase space $\mathcal{G} \times H$.}
    \label{fig:fig_04}
\end{figure}

\begin{figure}[H]
    \centering
    \includegraphics[scale=0.36]{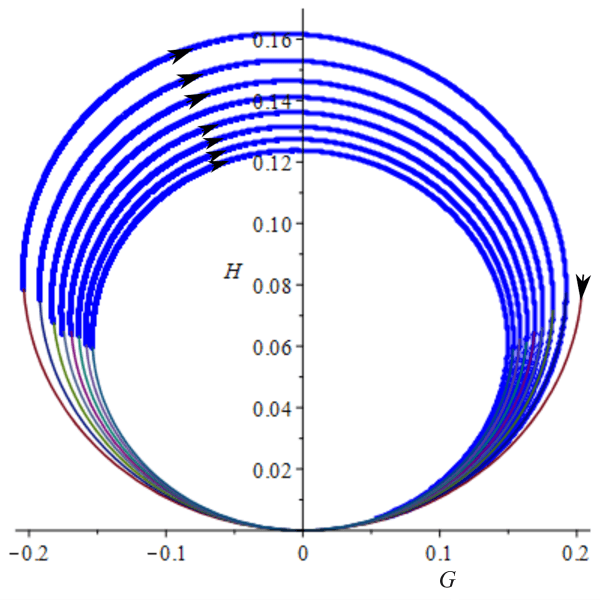}
    \caption{One example of an integral solution presenting an open cyclic behavior (vortex like) near the origin, each time passing through the origin with an smaller arch of travel, that is, the origin acts as an attraction pole.}
    \label{fig:fig_05}
\end{figure}

The last step, to understand the whole qualitative picture for the behavior of this simplified problem, is to determine how the numerical solutions given in \textit{figure \ref{fig:fig_01}} interact with the solutions near the origin of the phase space (\ref{eqn:14}) and (\ref{eqn:15}). We can take a direct approach, first consider the solution (\ref{eqn:14}) with an arbitrary value for $a$, because of its square root, both $\mathcal{G}$ and $H$ have two extreme values in this solution; so choosing one of the extreme values and considering them as initial conditions for the 2D dynamical system (\ref{eqn:10}) and (\ref{eqn:11}), we can numerically integrate to obtain an integral solution. Two things can happen to this integral solution, first it can grow away from the origin of the phase space (see \textit{figure \ref{fig:fig_01}}), or it can go back to the origin in two ways: it can go back to the original solution (\ref{eqn:14}), in which case the integral solution is cyclic closed; or it can return but for a different value for the constant $a$ in (\ref{eqn:14}), in which case the integral solution is cyclic open, or have a vortex like behavior. The result is, depending on the initial conditions chosen, the integral solution grows away from the origin or it cycles back in a vortex like behavior nearing the origin of the phase space, \textit{figure \ref{fig:fig_04}} shows some examples of the first result, and \textit{figure \ref{fig:fig_05}} shows an example of the second result.

\begin{figure}[H]
    \centering
    \includegraphics[scale=0.6]{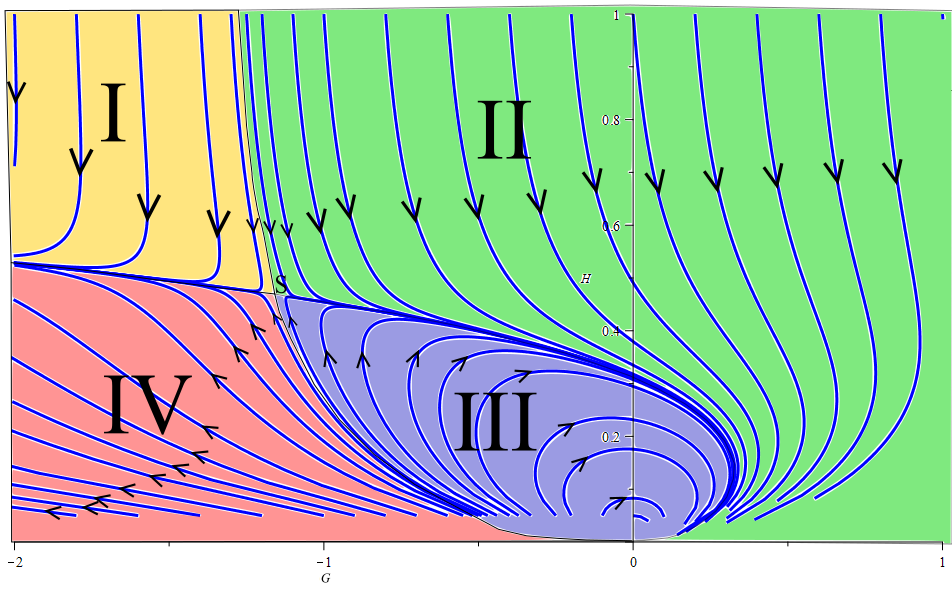}
    \caption{Global behavior of the 2D dynamical system(\ref{eqn:10}) and (\ref{eqn:11}) for $H$ positive. Quadrants (II) and (IV) are connected through the origin, and quadrant (III) have open cyclic solutions attracted to the origin.}
    \label{fig:fig_06}
\end{figure}

The \textit{figures \ref{fig:fig_01}, \ref{fig:fig_04} and \ref{fig:fig_05}} give us a summary of the global behavior for the 2D dynamical system (\ref{eqn:10}) and (\ref{eqn:11}). Focusing for $H$ positive, the phase space can be divided into four quadrants, see \textit{figure \ref{fig:fig_06}}. The first quadrant (I) gives us initial conditions where the Hubble parameter $H$ decreases to a minimum positive value and increases afterwards; the second quadrant (II) have initial conditions with integral solutions that converge to the origin of the phase space and are connected to the fourth quadrant (IV), then for initial conditions in (II) the Hubble parameter decreases to zero and increases afterwards in region (IV); and in region (III) the initial conditions result in integral solutions that are open cyclic around the origin of the phase space, where the origin acts as an attraction pole.

So now, having a global qualitative behavior for the empty space-time problem, this gives us a map of behavior to look for in the real case with energy-matter sources. The complete problem is given by the set of equations (\ref{eqn:6}) through (\ref{eqn:9}), for an empty space-time this set yields a 2D dynamical system, and adding matter-energy source yields a higher dimension dynamical system. Figures \ref{fig:fig_01} and \ref{fig:fig_06} provide a rough idea of how the integral solutions behave in these higher dimensional cases.

The behavior of the Hubble parameter, and consequently of the scale factor, is highly dependent on the initial conditions, as seen in \textit{figure \ref{fig:fig_06}}. On quadrant II it can describe a universe initially in an epoch of accelerated expansion, followed by a period of deceleration, and finally one last epoch of steady increase in acceleration and rate of expansion in region IV. Region I offers only the behavior of an ever increasing rate of expansion and acceleration of the system. Solutions on quadrants I and II/IV tend to converge to a single path in the phase space. While region III describe an empty Friedmann's universe in which the topological invariant $\mathcal{G}$ changes sign periodically, from (\ref{eqn:Q_Friedmann}) it implies that the acceleration of the scale factor $a(t)$ changes sign periodically, but the Hubble parameter $H$ is always positive or negative, in other words, it can describe an empty universe that is always expanding but periodically accelerating and decelerating, and moreover, the intensity of this acceleration and deceleration decreases over time. The points of equilibrium $\mathcal{G}_0$ and $H_0$ are saddle points, this implies that perturbations in these points may lead to two outcomes, either the system is attracted to the convergence of solutions of quadrants I and II/IV, or is attracted to the periodical solutions of quadrant III. The point $\mathcal{G}=0$ and $H=0$ is a special point of convergence and divergence of solutions, this means that small perturbations to the system near this origin may lead to radical differences in the evolution of the system, hinting to a possibly chaotic nature of behavior. In future work we will present a study when perturbations are applied to the system.  

From \eqref{eqn:0.1}, \eqref{eqn:0.2} and \eqref{eqn:0.5} we can infer that $\rho_z=3H^2$, and the geometrical equation of state relating the geometrical density and geometrical pressure is given by,
\begin{equation}
    p_z = \frac{\mathcal{G}}{4\rho_z}-\frac{\rho_z}{3}.
\end{equation}
Since the geometrical density $\rho_z$ is always positive, then the topological invariant $\mathcal{G}$ is the factor responsible for the sign of the effective pressure in the model. The geometrical density $\rho_z$ will be associated to a positive effective pressure when $\mathcal{G}$ satisfies the condition,
\begin{equation}
    \mathcal{G}>\frac{4}{3}\rho^2_z,
\end{equation}
and the negative pressure occurs in the condition,
\begin{equation}
    \mathcal{G}<\frac{4}{3}\rho^2_z.
\end{equation}
In particular, if $\mathcal{G}<0$ then $p_z<0$ is satisfied automatically.

\begin{figure}[H]
    \centering
    \includegraphics[scale=0.5]{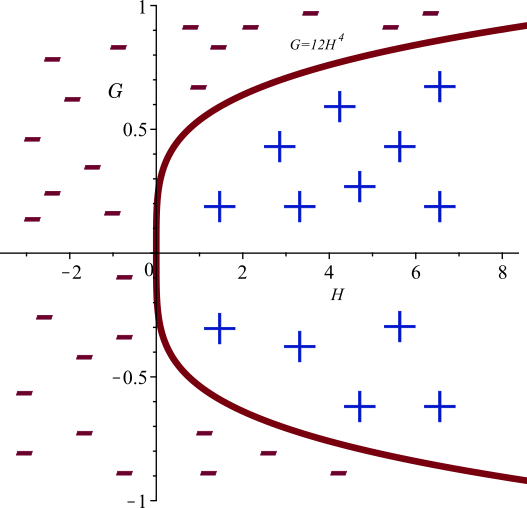}
    \caption{Regions in the phase space $\mathcal{G} \times H$ where the effective pressure is positive (region with blue plus), and negative pressure (region with negative red). The solid curve $\mathcal{G}=12H^4$ is where the pressure is zero.}
    \label{fig:fig_06b}
\end{figure}

In the phase space $\mathcal{G} \times H$, the positive and negative sign of the effective pressure is divided by the curve $\mathcal{G}=12H^4$ (see \textit{figure} \ref{fig:fig_06b}). If we compare with the global behavior map given in \textit{figure \ref{fig:fig_06}}, then, if $\mathcal{G}$ is in the cyclic region III, the effective pressure changes sign periodically, while in region II the pressure may change sign from negative to positive, and in region IV it becomes negative again, and finally, in region I the pressure is always negative.

\subsubsection{Friedmann's Universe with Cosmic Dust}

Let us consider now a Friedmann's universe with cosmic dust, in this case $p_m =0$. The set of differential equations are
\begin{align}
    \label{eqn:25}
    3H^2 = & \rho_z+\rho_m,\\
    \label{eqn:27}
    0 = & \dot{\rho_m}+3\rho_m H,\\
    \label{eqn:28}
    \mathcal{G}= & -24(\dot{H}+H^2)H^2.
\end{align}
This set of equations may be expressed as an autonomous 3D dynamical system in terms of the independent variables $\mathcal{G}$, $H$ and $\rho_m$ given by \eqref{eqn:0.4}, \eqref{eqn:0.5} and \eqref{eqn:0.6}, that is,

\begin{align}
    \label{eqn:29}
	\dot{\mathcal{G}}= & \frac{1}{24\sigma H^3}\left[3H^2-\sigma \frac{\mathcal{G}^2}{2}-\rho_m\right], \\
    \label{eqn:30}
    \dot{H}= & -\frac{\mathcal{G}}{24H^2}-H^2, \\
    \label{eqn:31}
    \dot{\rho_m}= & -3\rho_m H. \\
\end{align}

The equilibrium points of the system obtained by setting $\dot{\mathcal{G}}=0$, $\dot{H}=0$ and $\dot{\rho_m}=0$ are,

\begin{align}
    \mathcal{G}_o = & -\left(\frac{3}{2\sigma^2}\right)^{\frac{1}{3}}, \\
    H_o = & \pm\left(\frac{1}{96\sigma}\right)^{\frac{1}{6}}, \\
    \rho_{m_o} = & 0, \\
\end{align}
which are the same equilibrium points as in the case of an empty universe. The eigenvalues of the Jacobian matrix of the system, applied at the equilibrium points, have mixed signs, this implies they behave as saddle points for the integral solutions, just like in the empty space case. Besides these two equilibrium points, the solution $H=0$ satisfies equations \eqref{eqn:25} through \eqref{eqn:28}, with the conditions $\mathcal{G}=0$ and $\rho_m =0$, as it was shown in subsection IV.A.4, with $\rho_m =-\rho_z = \frac{1}{2}f(0) = 0$.

\begin{figure}[H]
     \centering
     \begin{subfigure}[b]{.49\textwidth}
         \centering
         \includegraphics[width=0.9\textwidth]{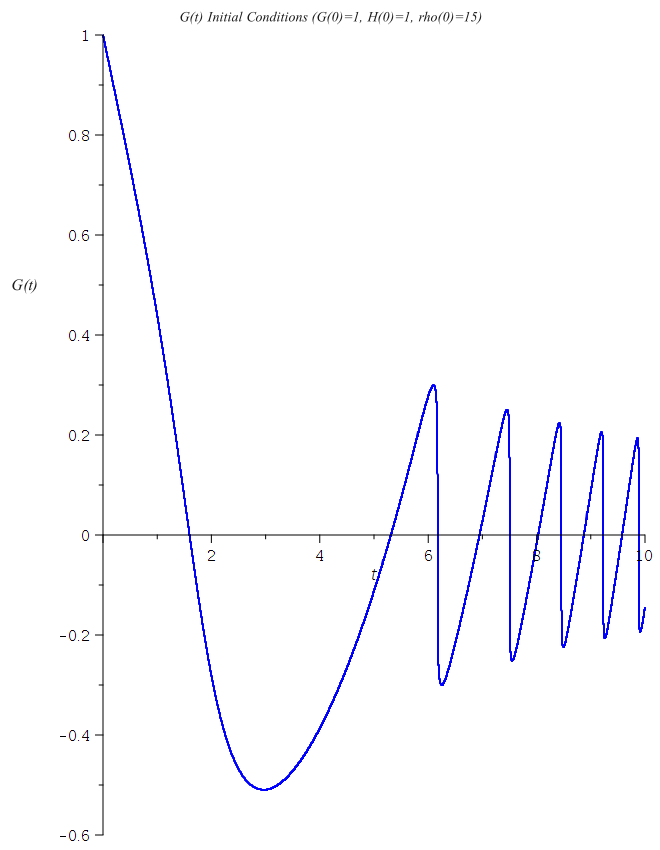}
         \caption{Plot of $\mathcal{G}(t)\times t$ for the integral solution in the phase space $[\mathcal{G}\times H\times \rho_m]$ with initial condition $\mathcal{G}(0)=1$, $H(0)=1$, $\rho(0)=15$.}
         \label{fig:fig_12}
     \end{subfigure}
     \hfill
     \begin{subfigure}[b]{.47\textwidth}
         \centering
         \includegraphics[width=0.9\textwidth]{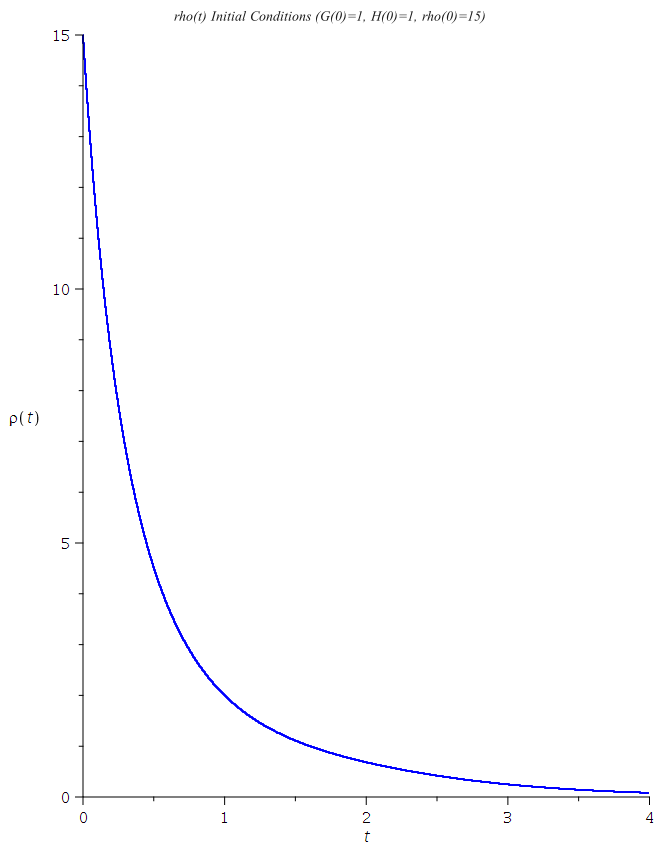}
         \caption{Plot of $\rho_m (t)\times t$ for the integral solution in the phase space $[\mathcal{G}\times H\times \rho_m]$ with initial condition $\mathcal{G}(0)=1$, $H(0)=1$, $\rho(0)=15$. Here $\rho_m$ satisfies $\rho_m \geq 0$.}
         \label{fig:fig_13}
     \end{subfigure}
     \caption{Numerical solution of the dynamical system in the phase space $\mathcal{G} \times H \times \rho_m$. Initial conditions $\mathcal{G}(0)=1$, $H(0)=1$, $\rho_m (0)=15$.}
\end{figure}

Numerical solutions to this system present behaviors like those observed in the empty space case, that is, for $H(t)$ and $\sigma$ positive, solutions in which the Hubble parameter decreases to a minimum positive or zero value and increases afterwards, or open cyclic behavior in the phase space. Since now we're dealing with a 3D dynamical system, a compilation of integral solutions in one 3D graph on the phase space becomes convoluted. A sample of two integral solutions, representing the two described behaviors, with given initial conditions, are given below:

In \textit{figures \ref{fig:fig_10}, \ref{fig:fig_11}, \ref{fig:fig_12} and \ref{fig:fig_13}} we have a case of integral solution with cyclic behavior. This shows that the cosmic dust has a dampening effect on the evolution of the system, the integral solution does not necessarily need to pass through the origin in the $\mathcal{G} \times H$ phase plane (as seen in \textit{figure \ref{fig:fig_10}}); the Hubble parameter $H$ starts with an initial value and then decreases periodically over time, like a ball bouncing from a surface a few times, approaching a stable value (as observed in \textit{figure \ref{fig:fig_11}}); the topological invariant $\mathcal{G}$ changes its sign a few times until a point in time when it approaches a stable value (as viewed in \textit{figure \ref{fig:fig_12}}), that is, it describes a Friedmann's universe that goes through epochs of accelerated and decelerated expansion, approaching a stable value with zero acceleration; and finally, the matter density decreases as expected.

\begin{figure}[H]
     \centering
     \begin{subfigure}[b]{.48\textwidth}
         \centering
         \includegraphics[width=0.64\textwidth]{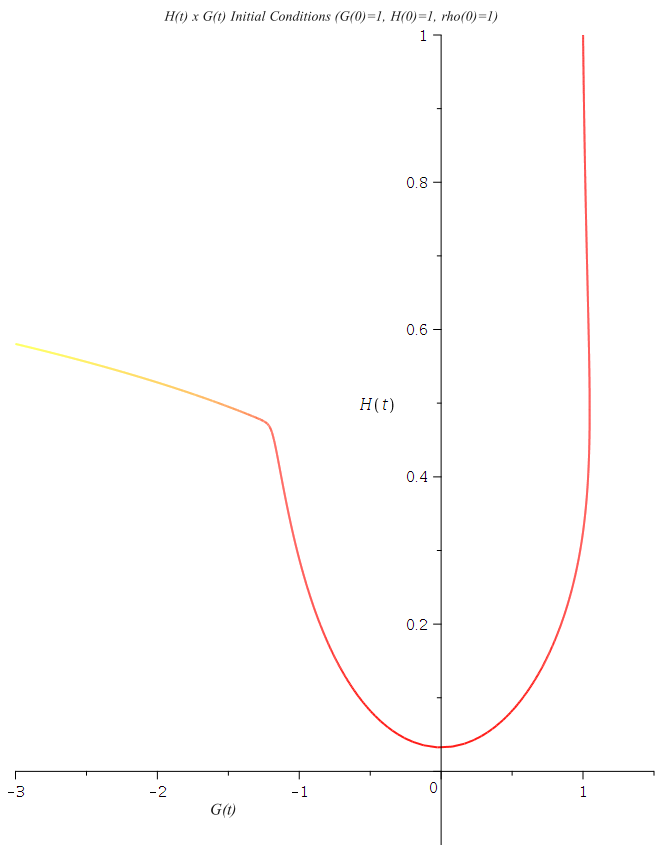}
         \caption{Plot of $\mathcal{G}(t)\times H(t)$ for the Integral solution in the phase space $[\mathcal{G}\times H\times \rho_m]$ with initial condition $\mathcal{G}(0)=1$, $H(0)=1$, $\rho(0)=1$.}
         \label{fig:fig_14}
     \end{subfigure}
     \hfill
     \begin{subfigure}[b]{.48\textwidth}
         \centering
         \includegraphics[width=0.64\textwidth]{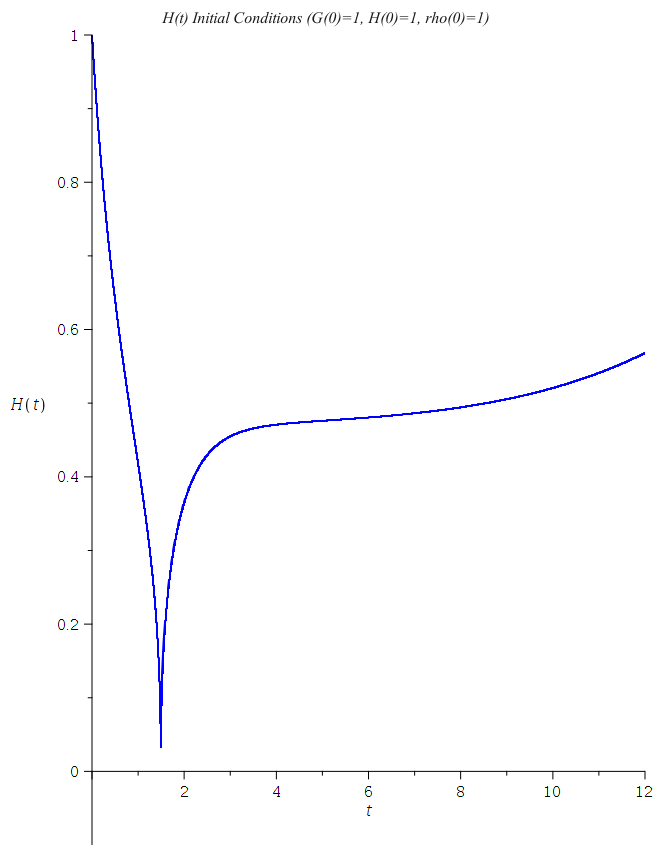}
         \caption{Plot of $H(t)\times t$ for the Integral solution in the phase space $[\mathcal{G}\times H\times \rho_m]$ with initial condition $\mathcal{G}(0)=1$, $H(0)=1$, $\rho(0)=1$. The $H(t)$ parameter has one minimum value.}
         \label{fig:fig_15}
     \end{subfigure}
     \caption{Numerical solution of the dynamical system in the phase space $\mathcal{G} \times H \times \rho_m$. Initial conditions $\mathcal{G}(0)=1$, $H(0)=1$, $\rho(0)=1$.}
\end{figure}

\begin{figure}[H]
     \centering
     \begin{subfigure}[b]{.48\textwidth}
         \centering
         \includegraphics[width=0.64\textwidth]{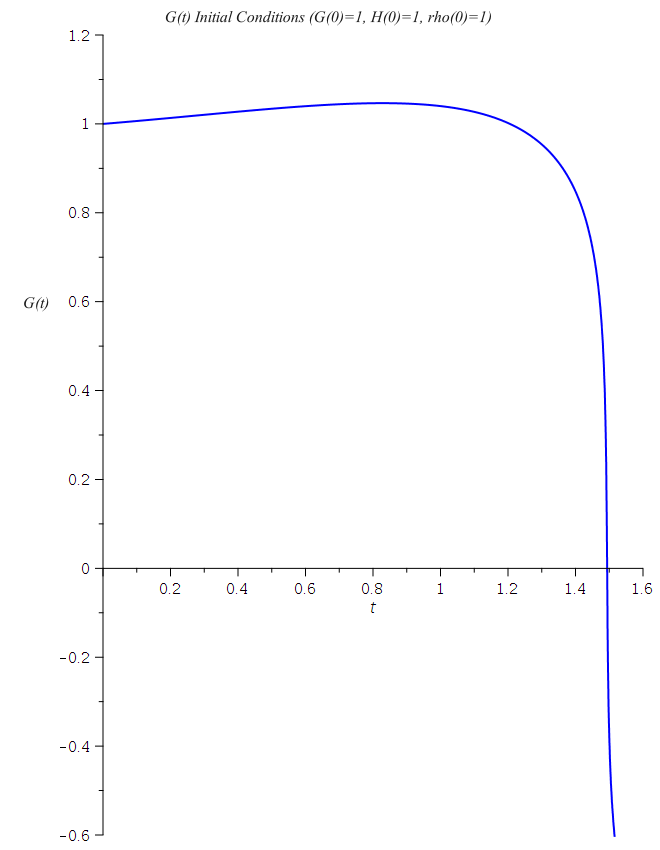}
         \caption{Plot of $\mathcal{G}(t)\times t$ for the Integral solution in the phase space $[\mathcal{G}\times H\times \rho_m]$ with initial condition $\mathcal{G}(0)=1$, $H(0)=1$, $\rho(0)=1$. $\mathcal{G}$ changes sign only one time.}
         \label{fig:fig_16}
     \end{subfigure}
     \hfill
     \begin{subfigure}[b]{.48\textwidth}
         \centering
         \includegraphics[width=0.64\textwidth]{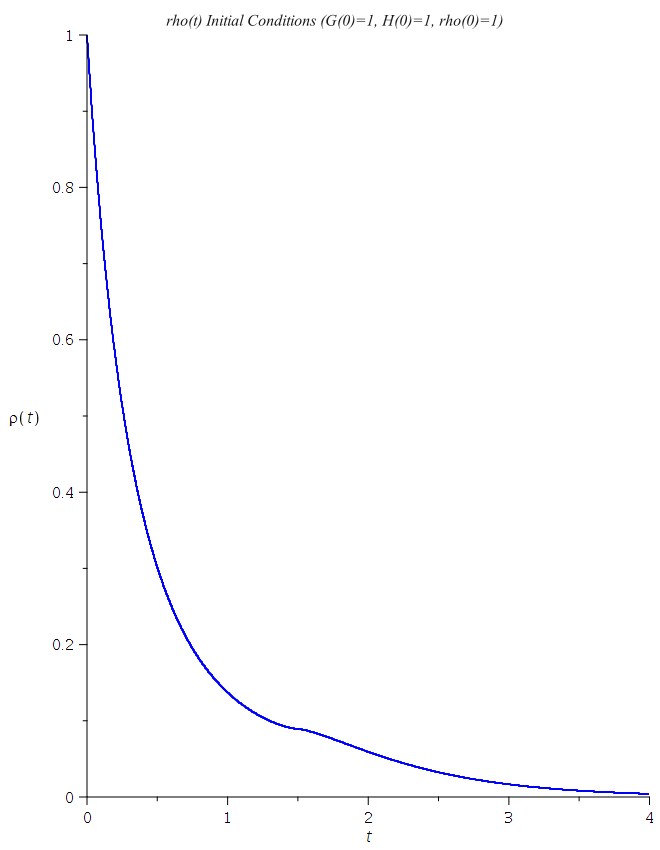}
         \caption{Plot of $\rho_m (t)\times t$ for the Integral solution in the phase space $[\mathcal{G}\times H\times \rho_m]$ with initial condition $\mathcal{G}(0)=1$, $H(0)=1$, $\rho(0)=1$.}
         \label{fig:fig_17}
     \end{subfigure}
     \caption{Numerical solution of the dynamical system in the phase space $\mathcal{G} \times H \times \rho_m$. Initial conditions $\mathcal{G}(0)=1$, $H(0)=1$, $\rho(0)=1$.}
\end{figure}

\textit{Figures \ref{fig:fig_14}, \ref{fig:fig_15}, \ref{fig:fig_16} and \ref{fig:fig_17}} show a case where the Hubble parameter has only one minimum positive value, and the topological invariant changes sign only one time.

One important qualitative result here is that, given a positive value for $\sigma$, and fixed initial conditions for $\mathcal{G}(0)$ and $H(0)=0$, the initial value of the cosmic dust density $\rho_m(0)$ will dictate which type of behavior the system will develop, that is, if the Hubble parameter will have only one minimum value, or have a cyclic behavior with epochs of acceleration and deceleration. Then $\rho_m$ plays an important role on the evolution of the system. 

\begin{figure}[H]
     \centering
     \begin{subfigure}[b]{.48\textwidth}
         \centering
         \includegraphics[width=\textwidth]{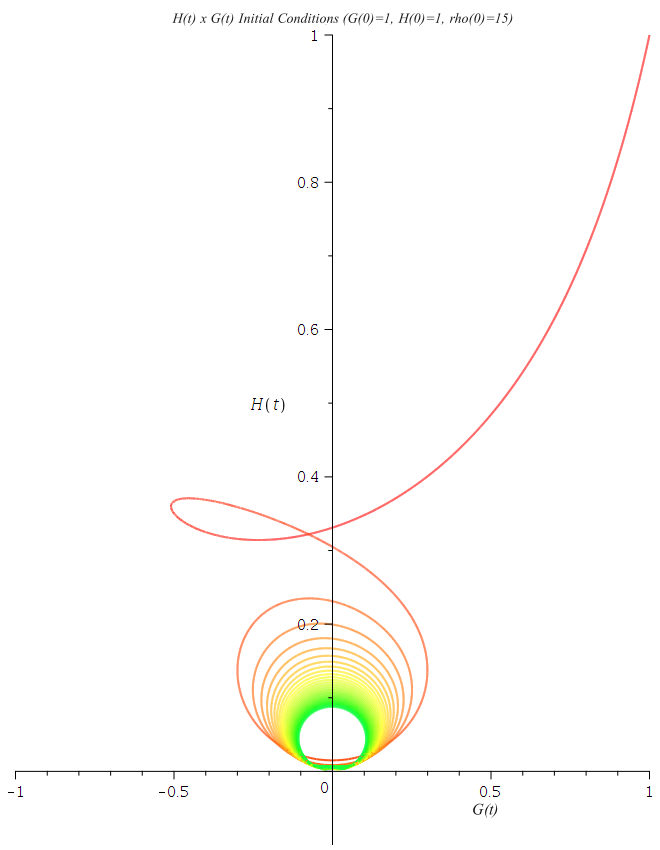}
         \caption{Plot of $\mathcal{G}(t)\times H(t)$ for the Integral solution in the phase space $[\mathcal{G}\times H\times \rho_m]$ with initial conditions $\mathcal{G}(0)=1$, $H(0)=1$, $\rho_m (0)=15$. Observe that this is a 3D phase space, so the path of the integral solution does not cross with itself but rather the path is flattened out in the plane $\mathcal{G}\times H$.}
         \label{fig:fig_10}
     \end{subfigure}
     \hfill
     \begin{subfigure}[b]{.46\textwidth}
         \centering
         \includegraphics[width=\textwidth]{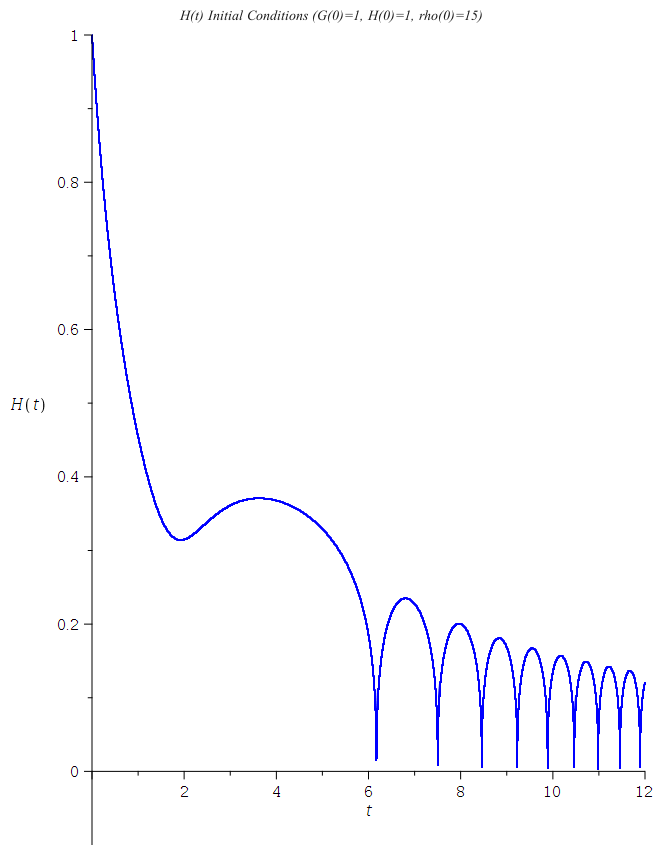}
         \caption{Plot of $H(t)\times t$ for the Integral solution in the phase space $[\mathcal{G}\times H\times \rho_m]$ with initial condition $\mathcal{G}(0)=1$, $H(0)=1$, $\rho_m (0)=15$. If you compare to the case with empty space-time (see figure \ref{fig:fig_05} for example), the cosmic dust has a dampening effect on the open cyclic behavior of the system.}
         \label{fig:fig_11}
     \end{subfigure}
     \caption{Numerical solution of the dynamical system in the phase space $\mathcal{G} \times H \times \rho_m$. Initial conditions $\mathcal{G}(0)=1$, $H(0)=1$, $\rho_m (0)=15$.}
\end{figure}

\vfill
\newpage

The dampening effect due to the cosmic dust in the system is a consequence of its direct action on $\dot{\mathcal{G}}$, as it can be seen in \eqref{eqn:29}. When the system is about to change from $\mathcal{G}$ positive to $\mathcal{G}$ negative, this implies $\dot{\mathcal{G}} < 0$, and we see from \eqref{eqn:29} and \eqref{eqn:30} that, the presence of the cosmic dust increases the negative value of $\dot{\mathcal{G}}$ but does not affect directly the $\dot{H}$. The consequence of this modification to the dynamical system is that, the cosmic dust accelerates the change of sign of the topological invariant from positive to negative before the Hubble parameter can reach zero. For example, consider the cyclic solution in \textit{figure \ref{fig:fig_10}}, when it is approaching the change of sign of $\mathcal{G}$ from positive to negative, this implies the condition over the Hubble parameter,
\begin{equation}\label{eqn:32}
    \dot{\mathcal{G}}<0 \rightarrow 3H^2 < \frac{\mathcal{G}^2}{2} + \rho_m,
\end{equation}
and when the topological invariant is exactly zero we have the specific condition,
\begin{equation}\label{eqn:33}
    \dot{\mathcal{G}}<0 \; and \; \mathcal{G}=0 \rightarrow 3H^2 < \rho_m.
\end{equation}
The above condition shows a mechanism for which the cosmic dust $\rho_m$ limits the maximum value that the Hubble parameter $H$ can have when the topological invariant $\mathcal{G}$ change sign from positive to negative, that is, when the model change from an epoch of decelerated expansion to an epoch of accelerated expansion. As the $\rho_m$ tends to zero (see \textit{figure \ref{fig:fig_13}}), the condition \eqref{eqn:33} implies that $H\rightarrow 0$, which is the behavior of the empty spacetime solution observed, for example, in \textit{figure \ref{fig:fig_05}}. 

\section{Conclusions}

The rank-3 tensor representation of a spin-2 field appears directly in the modified equations of $f(\mathcal{G})$ gravity as a covariant divergent. In a homogeneous and isotropic Friedmann's universe, the trivial case for $\mathcal{G}$ as a constant gives rise to specific conditions for $\rho_m\geq 0$ to be true, that is, depending on the sign of $\mathcal{G}$, the geometric density $\rho_z$ must satisfy certain conditions and the Hubble parameter $H(t)$ may only vary over delimited time intervals. In other words, in some cases the topological invariant $\mathcal{G}$ can only be constant over a finite interval of time in order for the density mass $\rho_m$ to be strictly positive, this gives one necessary, but not a sufficient tool to filter out non-physical solutions.

The first non-trivial case we can consider, $f(\mathcal{G})=\sigma G^2$, gives rise to different behaviors for the Friedmann's universe. In the empty space-time case, it can describe a universe that is always in accelerated expansion, or that is decelerated and accelerates at a later phase; or describe a universe that is expanding in phases of decreasing acceleration and deceleration. When cosmic dust is added to the system, its initial value will dictate how this universe will evolve, if it will expand with increasing dampened phases of acceleration and deceleration, or if it will expand with ever increasing acceleration. The Hubble parameter $H$ as a non zero constant is a viable solution for an empty space-time, while $H=0$ is a possible solution when a constant source matter $\rho_m$ is added.

The next step would to add a gas of photons to the system and study its influence. And to compare the numerical solutions with real observational data to constrain the value of $\sigma$ and verify if it is reasonable. In our dynamical systems $\mathcal{G}$ was unbounded, it would be worthwhile to bound $\mathcal{G}$ to a finite interval using a Born-Infeld type function, summing or multiplying the original function, and verify how this boundary for $\mathcal{G}$, and its variation, would affect the evolution of the system. Another point of interest would be to study perturbations to determine how stable and unstable the equilibrium points are, and in particular there is the origin of the phase space studied that may present chaotic behavior in the evolution of the system.

\section*{Acknowledgements}

M. Novello thanks Fundação de Amparo e Pesquisa do Estado do Rio de Janeiro (FAPERJ)  for a fellowship

Fábio H. M. dos Anjos thanks Conselho Nacional de Desenvolvimento Científico e Tecnológico (CNPq) fot the support

\bibliography{bibliografia}
\bibliographystyle{ieeetr}

\end{document}